\documentclass[11pt]{article}
\usepackage{graphicx}
\def\mydate{31 July 2004}

\def\ignore#1{{}}

\let\oldtheequation=\theequation
\def\doteqs#1{\setcounter{equation}{0}            
\def\theequation{{#1}.\oldtheequation}}
\newcounter{sxn}
\def\sx#1{\addtocounter{sxn}{1} \vskip 1.cm  \goodbreak
\noindent{\large\bf\leftline{\thesxn.~~#1}} \nobreak \vskip -.5cm}
\def\sxn#1{\sx{#1} \doteqs{\thesxn}}

\newcounter{axn}

\date{}

\newdimen\mybaselineskip
\newcommand{\beeq}{\begin{equation}}
\newcommand{\eneq}{\end{equation}}
\newcommand{\beqn}{\begin{eqnarray}}
\newcommand{\eeqn}{\end{eqnarray}}

\def\mybig{\displaystyle \strut }

\def\dd{\partial}
\def\la{\raise.16ex\hbox{$\langle$}\lower.16ex\hbox{}  }
\def\ra{\, \raise.16ex\hbox{$\rangle$}\lower.16ex\hbox{} }
\def\go{\rightarrow}

\def\onehalf{ \hbox{${1\over 2}$} }

\def\eff{{\rm eff}}
\def\GUT{{\rm GUT}}
\def\SUSY{{\rm SUSY}}

\def\psibar{ \psi \kern-.65em\raise.6em\hbox{$-$} \lower.6em\hbox{} }
\def\psibarb{ \psi \kern-.65em\raise.6em\hbox{$-$}  }

\def\myfrac#1#2{{\mybig #1\over \mybig #2}}

\begin{document}

\thispagestyle{empty}

\baselineskip=12pt

{\small \noindent \mydate \hfill OU-HET 477/2004}

\vspace*{3.cm}

\begin{center}  
{\LARGE \bf Dynamical Gauge-Higgs Unification\footnote{Contribution paper 
for ICHEP 2004.}} \\
\end{center}

\vskip .5cm 
\centerline{\Large (No. 12-0068)}

\baselineskip=14pt

\vspace{2cm}
\begin{center}
{\bf  Yutaka Hosotani}
\end{center}

\centerline{\small \it Department of Physics, Osaka University, Toyonaka, 
Osaka 560-0043, Japan}

\vspace{3cm}
\begin{abstract}
Gauge fields and both adjoint and fundamental Higgs fields are  unified in gauge theory defined on an orbifold.  It is shown how the Hosotani mechanism at the quantum level resolves the problem of the arbitrariness in boundary conditions imposed at the fixed points of the orbifold.  The role of adjoint Higgs fields in the standard GUT, which are extra-dimensional components of gauge fields in the current scheme, is taken by the Hosotani mechanism and additional dynamics governing 
the selection of equivalence classes
of boundary conditions.  The roles of fundamental Higgs fields, namely those of inducing the electroweak symmetry breaking and giving masses to quarks and leptons, are taken by the Hosotani mechanism and by extra twists in boundary conditions for matter. SUSY scenario nicely fits this scheme.  Explicit models are given for the gauge groups $U(3) \times U(3)$, $SU(5)$, and $SU(6)$ on the orbifolds $M^4 \times  (S^1/Z_2)$ 
and $M^4 \times (T^2/Z_2)$. 
\end{abstract}

\baselineskip=20pt plus 1pt minus 1pt

%%%%%%%%%%% 1 %%%%%%%%%%
\newpage

\sxn{Introduction }
Gauge theory in higher dimensions, particularly gauge theory on 
orbifolds,  
has been studied extensively in hoping to resolve the long-standing 
problems in grand unified theory (GUT) such as the gauge hierarchy problem,
the doublet-triplet splitting problem, and the
origin of gauge symmetry breaking.\cite{Antoniadis1}-\cite{Hall1}
One intriguing aspect is the  gauge-Higgs unification  in which Higgs bosons 
are regarded as  a part of extra-dimensional components of gauge 
fields.\cite{Manton1}-\cite{gaugeHiggs3}

When extra-dimensional space is not simply connected, dynamical gauge
symmetry breaking can occurs through the Hosotani 
mechanism, gauge symmetry breaking by the Wilson lines.\cite{YH1, YH2} 
Extra-dimensional 
components of  gauge
fields  (Wilson line phases) become  dynamical degrees of freedom, which
cannot be gauged away. They, in general circumstances, develop
nonvanishing  vacuum expectation values. 
Extra-dimensional components of  gauge fields 
act as  Higgs bosons at low energies.  Thus  gauge fields and Higgs
particles are  unified through  higher  dimensional gauge invariance. One does not
need to introduce extra Higgs fields to break the gauge symmetry.        
The gauge invariance also  protects Higgs fields from aquiring large masses by radiative corrections.

To construct realistic GUT or unified electroweak theory, one can choose 
 extra dimensions to be an orbifold.  By having an
orbifold in extra dimensions, one can accommodate chiral fermions in
four dimensions, and also rich patterns of gauge symmetry breaking. 
In this paper we  discuss gauge theory on $M^4 \times (S^1/Z_2)$ and 
$M^4 \times (T^2/Z_2)$.

\sxn{Gauge-Higgs unification}

The idea of unifying Higgs scalar fields with gauge fields was first proposed
by Manton and Fairlie\cite{Manton1}. Manton considered $SU(3)$ or $G_2$ 
gauge theory on $M^4 \times S^2$.  He, in ad hoc way, supposed that field strengths
on $S^2$ are nonvanishing in such a way that gauge symmetry breaks down to
the electroweak $SU(2)_L \times U(1)_Y$.  Extra-dimensional components of 
gauge fields of the broken part are the Weinberg-Salam Higgs fields.  
One of the serious problems in this senario is the fact that the configuration with 
nonvanishing field strengths has higher energy density than the trivial configration
with vanishing field strengths so that it will decay.  The stability is not
guaranteed even if the $S^2$ topology of the extra-dimnsional space is maintained
for other causes. 

There is a natural way of implementing the gauge-Higgs unification.  In 1983 
it was shown that in gauge theory defined on non-simply connected space,
dynamics of Wilson line phases can induce gauge symmetry breaking.  Particularly
it was proposed there that adjoint Higgs fields in GUT are extra-dimensional 
components of gauge fields. Dynamical symmetry breaking 
$SU(5) \go SU(3) \times SU(2) \times U(1)$ can take place at the quantum level
by the Hosotani mechanism.\cite{YH1}

Recently it has been found that in gauge theory defined on orbifolds 
boundary conditions at fixed points on orbifolds can implement 
gauge symmetry breaking.   It is subsequently pointed out that 
different sets of boundary conditions can be physically equivalent
through the Hosotani mechanism.  Consequently quantum treatment of Wilson
line phases becomes crucial to determine the physical symmetry of
the theory.\cite{HHHK}

Before going into the details, we stress that there are two types of
gauge-Higgs unification.

\vskip .3cm

\noindent {\bf (i) Gauge-adjoint-Higgs unification}

In most of grand unified theories (GUT), Higgs fields in the adjoint representation
are responsible for inducing gauge symmetry breaking down to the standard
model symmetry, $SU(3) \times SU(2) \times U(1)$.   The expectation value 
of such Higgs fields is typically of O($M_\GUT$). In higher dimensional
gauge theory extra-dimensional components of gauge fields can serve as
Higgs fields in the adjoint representation in four dimensions at low
energies. This is called gauge-adjoint-Higgs unification.  It was 
first introduced in ref.\ \cite{YH1}.

\vskip .3cm

\noindent {\bf (ii) Gauge-fundamental-Higgs unification}

Electroweak symmetry breaking is induced by Higgs fields in the fundamental 
representation.  In the Weinberg-Salam theory they are $SU(2)_L$ doublets.
In the $SU(5)$ GUT they are in the {\bf 5} representation.  
Higgs fields in the fundamental 
representation have another important role of giving fermions finite
masses. 

To unify a scalar field in the fundamental representation with gauge fields,
the gauge group has to be enlarged, as the scalar field need to become a part
of gauge fields.  In Manton's approach,\cite{Manton1} 
the gauge group is $SU(3)$ or $G_2$.
In GUT one can start with $SU(6)$ which breaks to $SU(3) \times SU(2) \times 
U(1)^2$.

\sxn{Gauge theory on non-simply connected manifolds and orbifolds}

If the space is non-simply connected, Wilson line phases become
physical degrees of freedom.  Although constant Wilson line phases
yield vanishing field strengths, they are dynamical and affect physics.
At the classical level Wilson line phases label degenerate vacua.
The degeneracy is lifted by quantum effects. The effective potential of 
Wilson line phases become non-trivial.   Wilson line phases are
non-Abelian Aharonov-Bohm phases.  If the effective potential is 
minimized at nontrivial values of Wilson line phases, then the 
rearrangement of gauge symmetry takes place. Spontaneous
gauge symmetry breaking or enhancement is achieved dynamically.

A class of orbifolds are obtained by dividing non-simply connected
manifolds by discrete symmetry.  Examples are $S^1/Z_2$ and $T^2/Z_n$.
In the course of this ``orbifolding'' there appear fixed points under
the discrete symmetry operation.  Theory requires additional boundary
conditions at those fixed points.  It gives us benefit of eliminating
some of light modes in various fields.  Chiral fermions naturally
appears at low energies.  Some of Wilson line phases drops out from
the spectrum, while the others survive.  The surviving Wilson line
phases can dynamically alter the boundary conditions at the fixed points
and the physical symmetry of the theory.

Let us take an example. First consider $SU(N)$ gauge theory 
on $M^4 \times T^n$.  $x^\mu$ ($\mu=0, \cdots, 3$) and $y^a$ $(a=1, \cdots, n)$
are coordinates of $M^4$ and $T^n$, respectively.  Loop translation along 
the $a$-th axis on $T^n$ gives
\beqn
T_a &:&  \vec{y}+\vec{l}_a ~\sim~ \vec{y} \cr
\noalign{\kern 5pt}
&& \vec{l}_a = (0, \cdots, 2\pi R_a, \cdots, 0) ~~~(a=1, 2, \cdots, n)~.
\label{translation1}
\eeqn
Although $(x, \vec y)$ and $(x, \vec y + \vec l_a)$ represent the same
point on $T^n$, the values of fields need not be the same. In general 
\beqn
&&\hskip -1cm 
A_M(x, \vec y + \vec l_a) =
U_a A_M(x, \vec y ) \, U_a^\dagger  ~~~, \cr
\noalign{\kern 5pt}
&&\hskip -1cm 
\psi(x, \vec y + \vec l_a) 
  =  \eta_a \, T[U_a] \psi(x, \vec y) ~~~, \cr
\noalign{\kern 5pt}
&&\hskip -1cm 
[U_a , U_b ] = 0 ~~~, ~~~ U_a \in SU(N) ~~~(a,b =1,\cdots, n)~.
\label{BC1}
\eeqn
$\eta_a$ is a $U(1)$ phase factor.  $T[U_a] \psi= U_a \psi$ or   
$U_a \psi U_a^\dagger$ for $\psi$ in the fundamental or adjoint representation,
respectively. The boundary condition (\ref{BC1}) guanrantees that the physics is 
the same at $(x, \vec y)$ and $(x, \vec y + \vec l_a)$.  The condition
$[U_a , U_b ] = 0$ is necessary to ensure $T_a T_b = T_b T_a$.
The theory is defined with a set of boundary conditions $\{ U_a, \eta_a \}$. 

Similar construction is done for gauge theory on orbifolds.  Take 
$M^4 \times (T^n/Z_2)$ as an example. $Z_2$ orbifolding gives
\beeq
Z_2 ~:~  - \vec{y} ~\sim ~\vec{y} ~~~.
\label{parity1}
\eneq
Applied on $T^n$, this parity operation allows a fixed point $z$ where
the relation $\vec{z}=-\vec{z}+\sum_a m_a\vec{l}_a$ ($m_a=$ an integer)
is satisfied.  There appear $2^n$ fixed points on $T^n$. Combining it  with
loop translations $T_a$ in (\ref{translation1}),  one finds that parity
around each fixed point is also a symmetry:
\beeq
Z_{2, j} ~:~  \vec z_j - \vec{y} ~\sim ~ \vec z_j + \vec{y} 
  \quad (j=0, \cdots, 2^n -1) ~~~.
\label{parity2}
\eneq
Accordingly fields must satisfy additional boundary conditions.  To be
definite, let spacetime be $M^4 \times (T^2/Z_2)$, in which case
$\vec z_0 =(0,0)$, $\vec z_1 =(\pi R_1,0)$, $\vec z_2 =(0,\pi R_2)$,
and $\vec z_3 =(\pi R_1, \pi R_2)$. 

Under $Z_{2,j}$ in (\ref{parity2})
\beqn
&&\hskip -1cm 
\pmatrix{A_\mu \cr A_{y^a} \cr} (x, \vec z_j - \vec y) =
P_j \pmatrix{A_\mu \cr - A_{y^a} \cr} (x, \vec z_j + \vec y) \, 
   P_i^\dagger ~ ~, \cr
\noalign{\kern 10pt}
&&\hskip -1cm 
\psi(x, \vec z_j - \vec y) = \eta_j' \, T[P_j] \, (i\Gamma^4\Gamma^5)
\psi(x, \vec z_j + \vec y) \qquad (\eta_j' = \pm 1) \cr
\noalign{\kern 5pt}
&&\hskip -1cm 
\qquad (a=1,2, \quad j=0, 1, 2,3) ~~.
\label{BC3}
\eeqn
Here $P_j = P_j^{-1} = P_j^\dagger \in SU(N)$.  Not all $U_a$'s and $P_j$'s
are independent.  On $T^2/Z_2$, only three of them are independent.  One can show
that
\beqn
&&\hskip -1cm 
U_a = P_a P_0 ~~~,~~~
P_3 = P_2 P_0 P_1 = P_1 P_0 P_2 ~~~, \cr
\noalign{\kern 5pt}
&&\hskip -1cm 
\eta_a = \eta_0' \eta_a' = \pm 1  \quad (a=1,2) ~. 
\label{BC4}
\eeqn 
Gauge theory on $M^4 \times (T^2/Z_2)$ is specified with a set of
boundary conditions $\{ P_j , \eta_j' ~;~ j=0, 1, 2 \}$. 
If fermions $\psi$ in (\ref{BC3}) are 6-D Weyl fermions, i.e. 
$\Gamma^7 \psi = +\psi$ or $-\psi$ where $\Gamma^7 = \Gamma^0 \cdots \Gamma^5$,
then the boundary condition (\ref{BC3}) makes 4D fermions chiral.

At a first look, the original gauge symmetry is broken by the boundary
conditions if $P_0$, $P_1$ and $P_2$ are not proportional to the identity
matrix. This part of the symmetry breaking is often called the orbifold
symmetry breaking in the literature.  As we see below, however, the physical
symmetry of the theory can be different from the symmetry of the boundary 
conditions, and different sets of boundary conditions can be equivalent
to each other.

\sxn{Wilson line phases and the Hosotani mechanism}

It is important to recognize that sets of boundary conditions form
equivalence classes.  Under a gauge transformation
\beeq
A_M'
=\Omega \bigg( A_M -{i\over g}\partial_M\bigg) \Omega^{\dagger}
\label{gaugeT1}
\end{equation}
$A_M'$ obeys a new set of boundary conditions $\{ P_j' , U_a' \}$ where
\beqn
&&\hskip -1cm
P_j' =\Omega(x, \vec z_j -\vec y) \, P_j 
   \,\Omega(x, \vec z_j +\vec y)^\dagger ~, \cr
\noalign{\kern 5pt}
&&\hskip -1cm
U_a' = \Omega(x, \vec y + \vec l_a)\, U_a\, \Omega(x, \vec y)^\dagger ~, \cr
\noalign{\kern 5pt}
&&\hskip 0. cm
\hbox{provided ~} \dd_M P_j' = \dd_M U_a' = 0 ~~.
\label{newBC1}
\eeqn 
The set $\{ P_j' \}$ can be different from the set $\{ P_j \}$.  
When the relations in (\ref{newBC1}) are satisfied, we write
\beeq
\{ P_j' \} ~\sim~ \{ P_j \} ~~.
\label{relation1}
\eneq
This relation is transitive, and therefore is an equivalence
relation.  Sets of boundary conditions form {\bf equivalence classes of boundary 
conditions} with respect to the equivalence 
relation (\ref{relation1}). \cite{YH2, HHHK,  HHK}

The equivalence relation (\ref{relation1}) indeed implies the equivalence of
physics as a result of dynamics of Wilson line phases.  Wilson line phases
are zero modes ($x$- and $\vec y$-independent modes) 
of extra-dimensional components of gauge fields which satisfy
\beqn
&&\hskip -1cm
A_{y_a} = \sum_{\alpha \in H_W} \onehalf A_{y_a}^\alpha
     \lambda^\alpha ~~~, ~~~
     [A_{y_a} , A_{y_b} ] = 0 ~~~,  ~~~ (a,b=1, \cdots, n) ~, \cr
\noalign{\kern 5pt}
&&\hskip -1cm
H_W = \Big\{ ~ \lambda^\alpha ~ ; 
 ~ \{ \lambda^\alpha , P_j \} = 0 \quad (j= 0, \cdots, 2^n - 1) ~ \Big\} ~.
\label{wilson1}
\eeqn
Consistency with the boundary condition (\ref{BC3}) requires 
$\lambda^\alpha$ in the sum to belong to $H_W$.  
Given the boundary conditions, these Wilson line phases cannot be 
gauged away.  They are physical degrees of freedom.  They label
degenerate classical vacua.  To put it differently, Wilson line phases
parametrize flat dirrections in the classical potential.
The values of $\la A_{y_a} \ra$ are determined, at the quantum level, 
 from the location of the 
absolute minimum of  the effective potential $V_\eff [ A_{y_a} ]$.

{\bf Physical symmetry} is determined in the combination of the 
boundary conditions $\{ P_j , \eta_j' \}$ and the expectation values of
the Wilson line phases $\la A_{y_a} \ra$.   Physical symmetry is, 
in general, different from the symmetry of the boundary conditions.
As a result of quantum dynamics gauge symmetry can be dynamically broken
by Wilson line phases.

This is called {\bf the Hosotani mechanism}.   The mechanism on non-simply 
connected manifolds was put forward in ref.\ \cite{YH1}.  The importance of 
equivalence classes of boundary conditions was clarified in ref.\ \cite{YH2}.
The detailed analysis of the Hosotani mechanism in gauge theory
on orbifolds was given in ref.\ \cite{HHHK}.  The mechanism is summarized
as follows. 

\begin{itemize}
\parskip=-2pt
\item[1.]
Wilson line phases, $\theta_W$,   are physical degrees of freedom 
and specify degenerate classical vacua.

\item[2.]
Quantum effects lift the degeneracy.  The effective potential for
the Wilson line phases $V_\eff[\theta_W]$ is nontrivial at the quantum 
level.  The global minimum of $V_\eff[\theta_W]$ determines the physical
vacuum.

\item[3.]
If $V_\eff[\theta_W]$ is minimized at nontrivial $\theta_W$, gauge
symmetry is spontaneously broken or enhanced.

\item[4.]
Gauge fields and adjoint Higgs fields (zero modes of $A_y$) are
unified.

\item[5.]
Adjoint Higgs fields acquire finite masses at the one loop level.
Finiteness of the masses is guaranteed by the gauge invariance.

\item[6.]
Physics is the same within each equivalence class of boundary conditions.
It does not depend on  sets of boundary conditions to start with
so long as they belong to the same equivalence class.

\item[7.]
Physical symmetry of theory is determined by matter content.

\end{itemize}

In the mechanism Higgs fields are naturally identified with 
extra-dimensional components of gauge fields.  The expectation 
values of Higgs fields are determined dynamically.  It is
{\bf dynamical gauge-Higgs unification}.

\sxn{$SU(N)$ gauge theory on $M^4 \times T^n$}

On a torus $T^n$ the boundary conditions are given by (\ref{BC1}),
denoted by $\{ U_a ~(a=1, \cdots, n) \}$.   Making use of the 
commutativity relations $U_a U_b = U_b U_a$, one can show that
\beeq
\{ ~ U_a ~ \} ~\sim ~ \{ ~ I ~ \} ~~.
\label{relation2}
\eneq
In other words, there is only one equivalence class.  Physics
does not depend on $\{ U_a ~(a=1, \cdots, n) \}$.  In particular,
in pure gauge theory the gauge symmetry remains unbroken even if
nontrivial $U_a \in SU(N)$ are imposed.

\sxn{$SU(5)$ GUT on $M^4 \times  (S^1/Z_2)$}

Kawamura pointed out that in $SU(5)$ gauge theory on 
$M^4 \times  (S^1/Z_2)$ with  the boundary conditions 
\beeq
{\rm BC}_1 ~:~~
P_0 = \pmatrix{1 \cr & 1\cr && 1 \cr &&&1 \cr &&&& 1\cr} ~~~,~~~
P_1 = \pmatrix{1 \cr & 1\cr &&1 \cr&&& -1\cr &&&& -1\cr} ~~~,
\label{BC5}
\eneq
the triple-doublet Higgs mass splitting problem can be 
naturally solved.\cite{Kawamura}
In his model there are no Wilson line phases surviving.  
$SU(5)$ symmetry is broken to $SU(3) \times SU(2) \times U(1)$
by boundary conditions, and there are no colored Higgs triplets
to begin with.

A question arises about the choice of boundary conditions to be
imposed.  Why do one need to choose BC$_1$?  This problem is 
called as the arbitrariness problem of boundary conditions.\cite{YH4}
 It is known that
in $SU(N)$ gauge theory on $M^4 \times  (S^1/Z_2)$, there are
$(N+1)^2$ equivalence classes of boundary conditions.\cite{HHK}

One can start with 
\beeq
{\rm BC}_2 ~:~~
P_0 = 
P_1 = \pmatrix{1 \cr & 1\cr &&1 \cr&&& -1\cr &&&& -1\cr} ~~~,
\label{BC6}
\eneq
or more generally
\beqn
{\rm BC}_3  &:&
P_0 =  \pmatrix{1 \cr & 1\cr &&1 \cr&&& -1\cr &&&& -1\cr} ~~~,\cr
\noalign{\kern 10pt}
&&
P_1 = \pmatrix{  \cos \alpha & 0& 0& i \sin \alpha & 0\cr
                   0&  \cos \beta & 0& 0& i \sin \beta \cr
                   0& 0& 1 & 0 & 0\cr
                  -i \sin \alpha & 0& 0 &   -\cos \alpha & 0\cr
                    0& -i \sin \beta & 0& 0& -\cos \beta \cr} ~~~.
\label{BC7}
\eeqn
BC$_2$ is a special case of BC$_3$ with $\alpha=\beta=0$.  The 
detailed analysis of the theory with BC$_3$ was given in ref.\
\cite{HHHK}.

Note first that BC$_2$ and BC$_3$ belong to the same equivalence
class:
\beeq
{\rm BC}_2 ~\sim ~ {\rm BC}_3 ~~~.
\label{relation3}
\eneq
Symmetry of boundary conditions, however, depends on $\alpha$ and
$\beta$:
\beeq
\hbox{symmetry of BC} ~=~ 
\cases{SU(3) \times SU(2) \times U(1) 
  &for $(\alpha, \beta) = (0,0)$\cr
  \noalign{\kern 3pt}
  SU(2) \times U(1)^3 
  &for $(\alpha, \beta) = (\pi ,0), (0, \pi)$\cr
  \noalign{\kern 3pt}
  SU(2)^2 \times U(1)^2
  &for $(\alpha, \beta) = (\pi ,\pi)$\cr
  \noalign{\kern 3pt}
 U(1)^3 
  &otherwise.\cr} 
  \label{symmetry3}
  \eneq
The Hosotani mechanism tells us that once matter content
in the theory is specified, physical symmetry is uniquely determined.
It is of great interest to know if $SU(3) \times SU(2) \times U(1)$
symmetry remains intact in supersymmetric $SU(5)$ theory. 

To determine the physical vacuum, one need to evaluate the 
effective potential for the Wilson line phases.\cite{Lim1, HHHK, Takenaga1, LeeHo}
  With the aid of
gauge invariance, it suffices to evaluate the effective potential
in the theory with any values of $(\alpha, \beta)$ in BC$_3$.  
Take $(\alpha, \beta) = (0,0)$, or BC$_2$. Wilson line phases are
the components of $A_y$ marked with $\star$ in 
\beeq
A_y = \pmatrix{&&& \star &\star\cr 
               &&& \star &\star\cr 
               &&& \star &\star\cr
      		\star &\star& \star \cr
            \star &\star& \star \cr} ~~.
\label{wilson2}
\eneq
Employing the residual $SU(3) \times SU(2) \times U(1)$ symmetry
of boundary conditions,    one can reduce it to
\beeq
2 g R A_y = \pmatrix{&&& a \cr   &&&&b\cr  &&&\cr
      		a\cr &b\cr} ~~.
\label{wilson3}
\eneq
$a$ and $b$ are phases with a normalized period 2.

\begin{figure}[hb]
\centering  \leavevmode
%\psfrag{a}{{\LARGE $a$}}
%\psfrag{b}{{\LARGE $b$}}
%\psfrag{V}{{\hskip -2.2cm \LARGE $V_\eff/2C$}}
\includegraphics[width=8.cm]{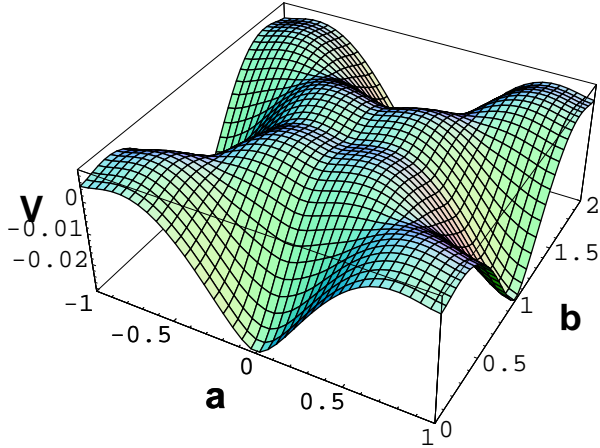}
\caption{$(32\pi^7 R^5/3) V_\eff(a,b)$ in (\ref{SUSYeffpot}) 
for $N_h=1$ and  $\beta=0.1$ is depicted.
For $N_h=1$ there are degenerate global minima at  
$(0,0)$ and $(1,1)$.}
\label{fig-Vsusy1}
\end{figure}

We consider supersymmetric $SU(5)$ model with $N_h$ Higgs scalar 
fields in {\bf 5} representation.  We suppose that quarks and leptons
are localized on the brane at one of the fixed points on $S^1/Z_2$.
 Supersymmetry breaking is introduced by
 Scherk-Schwarz $SU(2)_R$ twist.  The Scherk-Schwarz phase is 
 denoted by $\beta$.  Then the effective potential becomes
 \beqn
&&\hskip -1.5cm 
V_\eff (a,b)
= - {3\over 32 \pi^7 R^5} \sum_{n = 1}^{\infty}
{1 \over n^5} (1-\cos 2\pi n\beta) 
~ \bigg\{ 2(1- N_h)(\cos \pi na +\cos \pi nb) \cr
\noalign{\kern 5pt}
&&\hskip 4.0cm
 +4\cos \pi na  \cos \pi nb   +\cos 2\pi na +\cos 2\pi nb \bigg\} ~.
\label{SUSYeffpot}
\eeqn
In the minimal model, $N_h = 1$.  As displayed 
in fig.\ \ref{fig-Vsusy1},
$V_\eff (a,b)$ is minimized at $(a,b) = (0,0)$ and $(1,1)$.
Physical symmetry at $(a,b) = (0,0)$ is 
$SU(3) \times SU(2) \times U(1)$, whereas 
$SU(2) \times SU(2) \times U(1) \times U(1)$ at $(a,b) = (1,1)$.
In the minimal supersymmetric model these two phases are
degenerate.  For $N_h \ge 2$, $(a,b) = (1,1)$ is the global
minimum.  One sees that the standard model symmetry can be obtained
only for $N_h \le 1$.

In this model $M_\GUT \sim 1/R$.  Supersymmetry breaking scale
is given by $M_\SUSY \sim \beta /R \sim \beta M_\GUT$. 
Adjoint Higgs bosons ($A_y$ in $H_W$) acquire masses of 
$g_4 M_\SUSY$ where $g_4$ is the four-dimensional gauge 
coupling.

\sxn{$U(3)_S \times U(3)_W$ model on $M^4 \times (T^2/Z_2)$}

In the $SU(5)$ model described above, the Higgs fields in the 
fundamental representation are not unified.  To achieve the 
gauge-fundamental-Higgs unification one has to enlarge 
the gauge group such that fundamental Higgs fields in group $G$ 
can be identified with a part of gauge fields in the enlarged group
$\hat G$.  

The original proposal by Manton was along this line, but the
resultant low energy theory was far from the reality.
One interesting model was proposed by 
Antoniadis, Benakli and Quiros a few years ago.\cite{gaugeHiggs1}
They start with a product of two gauge groups 
$U(3)_S \times U(3)_W$ with gauge couplings $g_S$ and $g_W$. 
$U(3)_S$ is ``strong'' $U(3)$ which decomposes to 
color $SU(3)_c$ and $U(1)_3$.  $U(3)_W$ is ``weak'' $U(3)$ which decomposes to  weak $SU(3)_W$ and $U(1)_2$.  The theory
is defined on $M^4 \times (T^2/Z_2)$.  Boundary conditions at
fixed points of $T^2/Z_2$ are imposed in the following manner.
For the $U(3)_S$ group, all $P_0$, $P_1$ and $P_2$ are taken to
be identity matrix. For $U(3)_W$ one takes
\beeq
P_0 = P_1 = P_2 = 
\pmatrix{-1 \cr &-1\cr &&+1\cr} ~~~.
\label{BC8}
\eneq
The boundary condition (\ref{BC8}) breaks $SU(3)_W$ to
$SU(2)_L \times U(1)_1$ at the classical level.  
There are three $U(1)$'s left over.

Fermions obey boundary condition in (\ref{BC3}).  Let 
$(n_S, n_W)^\sigma$ stand for a fermion in the $n_S$ ($n_W$)
representation of $U(3)_S$ ($U(3)_W$)  with 6D-Weyl
eigenvalue $\Gamma^7 = \sigma$.   Three generations of
leptons are assigned as follows.  Leptons are
\beeq
L_{1,2,3} = (1 , 3)^+ ~~~:~~~
\pmatrix{{\nu_L} \cr {e_L} \cr \tilde e_L \cr} ~,~
\pmatrix{\tilde \nu_R \cr \tilde e_R \cr  {e_R} \cr}~~ \hbox{etc}.
\label{lepton1}
\eneq
Similarly, for right-handed down quarks we have
\beeq
D^c_{1,2,3} = (\bar 3 , 1)^+ ~~~:~~~
{d^c_L}  ~~,~~ {\tilde d^c_R} ~~ \hbox{etc}.
\label{quark1}
\eneq
For other quarks, each generation has its own assignment:
\beqn
Q_{1}  = (3 , \bar 3)^+ 
&&
\pmatrix{{u_L} \cr {d_L} \cr \tilde u_L \cr} ~,~
\pmatrix{\tilde u_R \cr \tilde d_R \cr  {u_R} \cr} \cr
\noalign{\kern 10pt}
Q_{2}  = (3 , \bar 3)^-
&&
\pmatrix{{c_L} \cr {s_L} \cr \tilde c_L \cr} ~,~
\pmatrix{\tilde c_R \cr \tilde s_R \cr  {c_R} \cr} \cr
\noalign{\kern 10pt}
Q_{3}  = (\bar 3 ,  3)^-
&&
\pmatrix{\tilde t_L^c \cr \tilde b_L^c \cr {t_L^c} \cr} ~,~
\pmatrix{{t_R^c} \cr {b_R^c} \cr \tilde t_R^c \cr} ~.
\label{quark2}
\eeqn
Due to the boundary conditions either $SU(2)_L$ doublet
part or singlet part has zero modes.  In (\ref{lepton1})-(\ref{quark2}), 
fields with tilde $\tilde {~}$ do not have zero modes.

With these assignments of fermions only one combination of
three $U(1)$ gauge groups remains anomaly free, which is
identified with weak hypercharge $U(1)_Y$.  Gauge bosons 
corresponding to the other two combinations of three $U(1)$ 
gauge groups become massive by the Green-Schwarz mechanism.
Hence, the remaining symmetry at this level is
$SU(3)_c \times SU(2)_L \times U(1)_Y$.

\def\myb{{\vphantom{\myfrac{1}{2}}}}
There are Wilson line phases in the $SU(3)_W$ group.  They 
are 
\beeq
A_{y_1} = \pmatrix{&& \star\cr && \star\cr \star&\star\cr} 
=\pmatrix{~ & \myb \Phi_1 \cr
         ~~\Phi_1^\dagger ~ &  \cr}
         ~~,~~
 A_{y_2} 
=\pmatrix{~ & \myb \Phi_2 \cr
         ~~\Phi_2^\dagger ~ &  \cr} ~~.
\label{wilson4}
\eneq
$\Phi_1$ and $\Phi_2$ are $SU(2)_L$ doublets.  The resultant
theory is the Weinberg-Salam theory with two Higgs doublets.
The classical potential for the Higgs fields results from
the $F_{y_1 y_2}^2$ part of the gauge field action:
\beeq
V_{\rm tree}(\Phi_1, \Phi_2) =  g_W^2 
\Big\{ 
 \Phi_1^\dagger \Phi_1^{} \cdot \Phi_2^\dagger \Phi_2^{} 
+ \Phi_2^\dagger \Phi_1^{} \cdot \Phi_1^\dagger \Phi_2^{}
-(\Phi_2^\dagger \Phi_1^{})^2 - (\Phi_1^\dagger \Phi_2^{})^2 
\Big\} ~~.
\label{tree1}
\eneq
There is no quadratic term.  The potential (\ref{tree1}) is
positive definite and  has flat directions.  The potential
vanishes if $\Phi_1$ and $\Phi_2$ are proportional to each other
with a real proportionality constant.

To determine if the electroweak symmetry is dynamically broken, 
one need to evaluate quantum corrections to the effective potential
of $\Phi_1$ and $\Phi_2$.  The detailed analysis is given 
 in ref.\ \cite{HNT2}.  The effective potential in the flat 
 directions is obtained, without loss of generality, for 
 the configuration 
 \beeq
 2 g_W R_1 A_{y_1} = \pmatrix{&&0\cr &&a\cr 0&a\cr} ~~,~~
 2 g_W R_2 A_{y_2} = \pmatrix{&&0\cr &&b\cr 0&b\cr} ~~,
 \label{wilson5}
 \eneq
where $a$ and $b$ are real.
Our task is  to find $V_\eff(a,b)$ and thereby determine 
the physical vacuum. 

Depending on the location of the global minimum of $V_\eff(a,b)$,
the physical symmetry varies.  It is given by
\beeq
(a,b) = \cases{
(0,0)  &$SU(2)_L \times U(1)_Y$\cr
\noalign{\kern 5pt}
(0,1), (1,0), (1,1) ~~~&$U(1)_{EM} \times U(1)_Z$\cr
\noalign{\kern 5pt}
\hbox{otherwise} &$U(1)_{EM}$.\cr}
\label{symmetry5}
\eneq
For generic values of $(a,b)$, electroweak symmetry breaking
takes place.  The Weinberg angle is given by
\beeq
\sin^2 \theta_W
= {1\over 4 + \myfrac{2 g_W^2}{3 g_S^2}} ~~,
\label{Wangle}
\eneq 
which can be very close to the observed value.  The deviation 
from the value $0.25$ is brought by a small ratio $g_W/g_S$.
We note that
in the $SU(3)_c \times SU(3)_W$ model the Weinberg angle
turns out too large.\cite{gaugeHiggs3}

The evaluation of $V_\eff(a,b)$ is straightforward.  
A general method of computations on $T^2/Z_2$ has been described 
in ref.\ \cite{HNT1}. In the 
non-supersymmetric model the matter content is given by
gauge fields (including ghosts) and fermions summarized in
(\ref{lepton1})-(\ref{quark2}).  Only gauge fields in $SU(3)_W$
give contributions having the ($a,b$) dependence.  The result is
\beqn
&&\hskip -1cm
V_\eff(a,b) = 
4 \bigg\{ I(0,0) + 2 \cdot I\Big( {a\over 2}, {b\over 2} \Big) 
    + I(a,b) \bigg\} 
- 3 \bigg\{ 14 \cdot I(0,0) 
   + 16 \cdot I\Big( {a\over 2}, {b\over 2} \Big) \bigg\} \cr
\noalign{\kern 10pt}
&&\hskip .5cm
= - 40 \cdot I\Big( {a\over 2}, {b\over 2} \Big) 
    + 4 \cdot I(a,b) + \hbox{const.} 
\label{Veff2}
\eeqn
where
\beqn
&&\hskip -1cm
I(a,b) = - {1\over 16\pi^2} \bigg\{
{1\over R_1^6} \sum_{n=1}^\infty {\cos 2\pi n a \over n^6}
 +  {1\over R_2^6} \sum_{m=1}^\infty {\cos 2\pi m b \over m^6} \cr
 \noalign{\kern 10pt}
 &&\hskip 3cm 
 + \sum_{n=1}^\infty  \sum_{m=1}^\infty 
 {2 \cos 2\pi n a \cos 2\pi m b \over (n^2 R_1^2 +  m^2 R_2^2 )^3}
\bigg\} ~~.
\label{Veff3}
\eeqn
In the first equality in (\ref{Veff2}), the first and second terms
represents contributions from gauge fields 
and fermions, respectively.

\vskip -.2cm
\begin{figure}[h]
\centering 
%\hskip 2.5cm \leavevmode 
\includegraphics[width=7.cm]{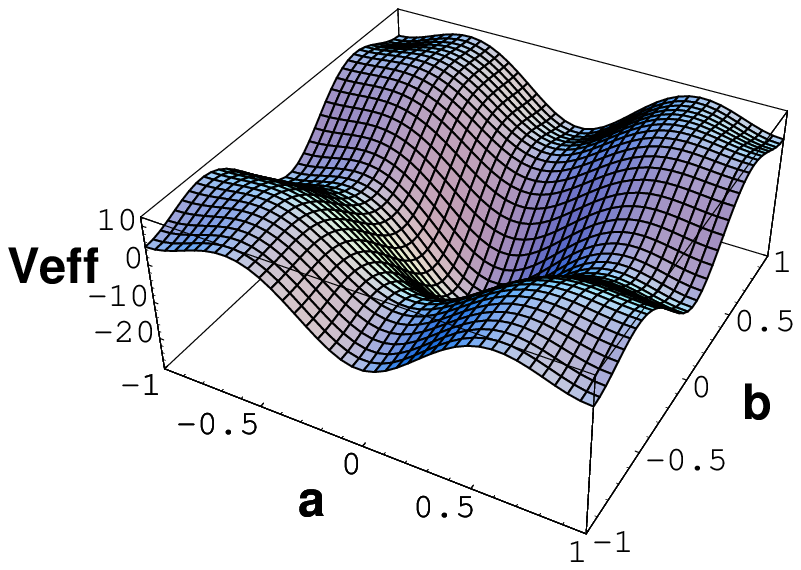}
\hskip 1cm 
\includegraphics[width=7.cm]{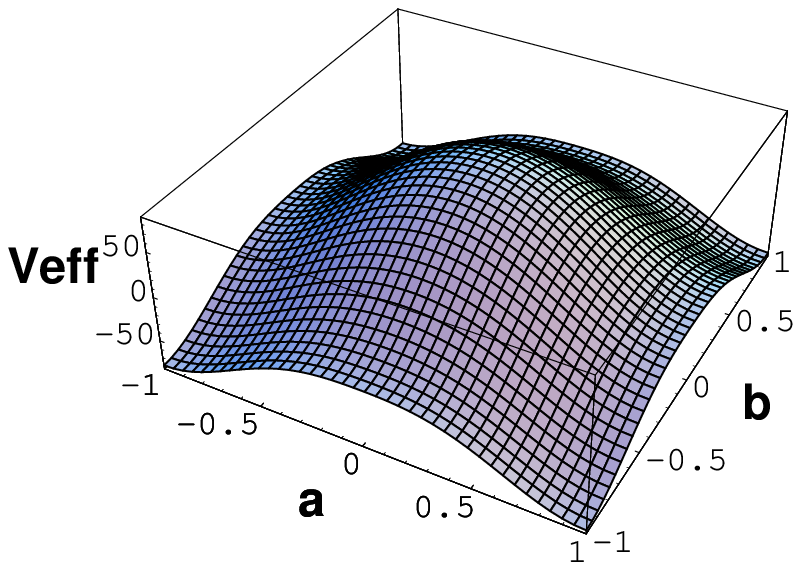}\\
{(a) In pure gauge theory. \hskip 4cm (b) With fermions.}
\caption{$V_\eff(a,b)$ in the $U(3)_S \times U(3)_W$ model.}
\label{fig-V-T2}
\end{figure}

If there were no fermions, $V_\eff(a,b)$ has the global minimum at
$(a,b) = (0,0)$ so that $SU(2)_L \times U(1)_Y$ symmetry is
unbroken.  In the presence of fermions, the point $(a,b) = (0,0)$
becomes unstable.  The effective potential (\ref{Veff2}) is 
displayed in fig.\ \ref{fig-V-T2}.  The global minimum is located at
$(a,b) = (1,1)$, which corresponds to the $U(1)_{EM} \times
U(1)_Z$ symmetry.  Although the $SU(2)_L$ symmetry is partially
 broken and $W$ bosons acquire masses, $Z$ bosons remain massless.
 
This result is not what we hoped to obtain.  We would like to have
a model in which the global minimum of the effective potential
is located at non-integral values of $(a,b)$.  As Antoniadis et al.\
mentioned in ref.\ \cite{gaugeHiggs1}, more general symmetry
breaking may occur if one considers a two-torus of general
parallelogram.  (In this section a rectangular torus has been
 considered.)  More promissing is to incorporate additional
 fermions, for instance, in the adjoint representation.  One can show
 that such modification indeed yields the global minimum
 at a generic point.\cite{HNT2}

\sxn{$SU(6)$ model on $M^4 \times (S^1/Z_2)$}

Gauge-fundamental-Higgs unification can be realized in the 
framework of GUT as well.  To illustrate it, let us consider 
$SU(6)$ gauge theory on $M^4 \times (S^1/Z_2)$.\cite{gaugeHiggs3}
We take boundary conditions to be
\beeq
P_0 = \pmatrix{1\cr &1\cr &&1\cr &&&1\cr &&&&-1\cr &&&&&-1\cr} ~~,~~
P_1 = \pmatrix{1\cr &-1\cr &&-1\cr &&&-1\cr &&&&-1\cr &&&&&-1\cr} ~~.
\label{BC9}
\eneq
Symmetry of boundary conditions is 
$SU(3) \times SU(2) \times U(1)^2$.  Wilson line phases are
\beeq
A_y = \pmatrix{0 &0 &0 &0 &\star &\star\cr
       0\cr 0\cr 0\cr \star\cr \star\cr}
       = \pmatrix{0 &0 &0 &0&~\Phi^\dagger ~~\cr
         0\cr 0\cr0 \cr \myb \Phi \cr} ~~.
         \label{wilson6}
\eneq
They serve as a Higgs doublet.  Electroweak symmetry breaking 
is induced if $\Phi$ dynamically develops an expectation value:
\beeq
2 gR ~ \la \Phi \ra = \pmatrix{0\cr a\cr} ~~.
\label{wilson7}
\eneq

The effective potential $V_\eff(a)$ depends on the matter content.
On $M^4 \times (S^1/Z_2)$ fermions satisfy
\beeq
\psi(x, z_j -  y) = \eta_j' \, T[P_j] \, \Gamma^5
\psi(x, z_j + y) \qquad (\eta_j' = \pm 1 ~,~ j= 0, 1) ~~.
\label{BC10}
\eneq
Here $z_0=0$ and $z_1= \pi R$.  Let $N_a^{(+)}$ ($N_f^{(-)}$)
be the number of fermions in the adjoint (fundamental)
representation with $\eta_0' \eta_1' = +1$ $(-1)$.  Then
\beqn
&&\hskip -0cm
V_\eff(a) = {3\over 64 \pi^7 R^5} \sum_{n=1}^\infty {1\over n^5}
\bigg\{ \Big( - {3\over 2} + 2 N_a^{(+)} \Big) \cos 2\pi n a \cr
\noalign{\kern 10pt}
&&\hskip 1cm
+ \Big( -3 + 4 N_a^{(+)} \Big) \cos \pi n a 
+ \Big(-9 + 12 N_a^{(+)} + 2 N_f^{(-)} \Big)
    \cos \pi n (a-1) \bigg\} ~~.
\label{Veff4}
\eeqn
When $N_a^{(+)} = N_f^{(-)} =2$, the global minimum is located at
$a=0.072$.  From the $W$ boson mass it follows that 
$a/g_4 R \sim 246 \,$GeV. 
The mass of the neutral Higgs is found to be
\beeq
m_H \sim {0.038 g_4 \over R} \sim 130 \, g_4^2 \, \hbox{GeV}~.
\label{mass2}
\eneq
In this senario $1/R$ is at a TeV scale.

The point of this example is to show that it is possible to 
have a small value for $a$ at the minimum, once one 
introduces additional  fermions.

\sxn{Summary}

We have shown in this paper that dynamical gauge-Higgs unification 
is achieved in higher dimensional gauge theory.  Higgs fields are
identified with Wilson line phases in gauge theory.  Dynamical
symmetry breaking is induced by the Hosotani mechanism.

Boundary conditions which appear in gauge theory on non-simply
connected manifolds or orbifolds are classified with equivalence
relations.  In each equivalence class of boundary conditions
physics is the same, as a consequence of quantum dynamics
of Wilson line phases.  

We have shown that both GUT symmetry breaking and electroweak
symmetry breaking can be induced in the present approach.  One of
the remaining problems is the origin of fermion masses.
Fermion masses brought by the Hosotani mechanism 
are flavor-independent.  They depend only on the representation 
of the group which fermions belong to.  There are other origins
for fermion masses.  There can be additional interactions
localized on the boundary brane.  We point out that there is a
natural origin of fermion masses on $T^n/Z_2$, namely
$T^n$ twists for $Z_2$ doublets.   In the case of fermions
on $M^4 \times (T^2/Z_2)$, we prepare a pair of fermion fields,
$(\psi, \hat \psi)$, and impose, instead of (\ref{BC1}) and 
(\ref{BC4}), 
\beqn
&&\hskip -1cm 
\pmatrix{\psi \cr \hat \psi \cr} (x,  - \vec y) 
= \eta_0' \, T[P_0] \, (i\Gamma^4\Gamma^5)
\pmatrix{\psi \cr -\hat \psi \cr} (x,  + \vec y)  \cr
\noalign{\kern 10pt}
&&\hskip -1cm 
\pmatrix{\psi \cr \hat \psi \cr} (x, \vec y + \vec l_a) 
  =  \eta_a \, T[U_a]
  \pmatrix{ \cos \gamma_a & -\sin \gamma_a \cr
           \sin \gamma_a & \cos \gamma_a \cr}
   \pmatrix{\psi \cr \hat \psi \cr} (x, \vec y) ~~~.
\label{BC11}
\eeqn
This is similar to the Scherk-Schwarz SUSY breaking.  If
twist parameters $\gamma_a$ are small, then the spectrum of light
particles at low energies does not change, but light fermions
acquire additional small masses of $O(\gamma_a/R)$.  

Finally we add a comment on the Higgsless model of electroweak
interactions recently proposed.\cite{Higgsless}
  The Higgsless model is very similar  to Kawamura's
model of $SU(5)$ gauge theory on 
$M^4 \times (S^1/Z_2)$.\cite{Kawamura}  In 
Kawamura's model colored triplet Higgs fields are absent due to
the boundary conditions.  In the Higgsless model boundary conditins
are designed such that Higgs doublet fields are absent.  In this
sense the Higgsless model also belongs to the category of models
examined in this paper. 

%%%%%%%%%%%%%%%%%%%%%%%%%%%%%%

% A useful Journal macro
\def\jnl#1#2#3#4{{#1}{\bf #2} (#4) #3}

\def\Zphys{{\em Z.\ Phys.} }
\def\jssc{{\em J.\ Solid State Chem.\ }}
\def\jpsJ{{\em J.\ Phys.\ Soc.\ Japan }}
\def\ptps{{\em Prog.\ Theoret.\ Phys.\ Suppl.\ }}
\def\PTP{{\em Prog.\ Theoret.\ Phys.\  }}

\def\JMP{{\em J. Math.\ Phys.} }
\def\NPB{{\em Nucl.\ Phys.} B}
\def\NP{{\em Nucl.\ Phys.} }
\def\PLB{{\em Phys.\ Lett.} B}
\def\PL{{\em Phys.\ Lett.} }
\def\PRL{\em Phys.\ Rev.\ Lett. }
\def\PRB{{\em Phys.\ Rev.} B}
\def\PRD{{\em Phys.\ Rev.} D}
\def\PR{{\em Phys.\ Rev.} }
\def\PRe{{\em Phys.\ Rep.} }
\def\AP{{\em Ann.\ Phys.\ (N.Y.)} }
\def\RMP{{\
em Rev.\ Mod.\ Phys.} }
\def\ZPC{{\em Z.\ Phys.} C}
\def\SCI{\em Science}
\def\CMP{\em Comm.\ Math.\ Phys. }
\def\MPLA{{\em Mod.\ Phys.\ Lett.} A}
\def\IJMPB{{\em Int.\ J.\ Mod.\ Phys.} B}
\def\cmp{{\em Com.\ Math.\ Phys.}}
\def\JPA{{\em J.\  Phys.} A}
\def\JPG{{\em J.\  Phys.} G}
\def\CQG{\em Class.\ Quant.\ Grav. }
\def\ATMP{{\em Adv.\ Theoret.\ Math.\ Phys.} }
\def\ibid{{\em ibid.} }
\vskip 1cm

\leftline{\bf References}

\renewenvironment{thebibliography}[1]
        {\begin{list}{[$\,$\arabic{enumi}$\,$]}  
% {\arabic{enumi}.}
        {\usecounter{enumi}\setlength{\parsep}{0pt}
         \setlength{\itemsep}{0pt}  \renewcommand{\baselinestretch}{1.2}
         \settowidth
        {\labelwidth}{#1 ~ ~}\sloppy}}{\end{list}}

%%%%%%%%%%%%%%%%%%

\end{document}